\documentclass{ws-ijmpcs}

\begin{document}

\markboth{K. MALTMAN}
{Lattice Input on the Inclusive $\tau$ Decay $V_{us}$ Puzzle}

%
\catchline{}{}{}{}{}
%

\title{Lattice Input on the Inclusive $\tau$ Decay $V_{us}$ Puzzle}

\author{P. A. BOYLE, L. DEL DEBBIO and R.J. HUDSPITH}
\address{Physics and Astronomy, University of Edinburgh\\
Edinburgh EH9 3JZ, UK}

\author{N. GARRON}
\address{School of Mathematics, Trinity College\\
Dublin 2, Ireland}

\author{E. KERRANE}
\address{Instituto de F\`isica T\`eorica UAM/CSIC, Universidad
Aut\`onoma de Madrid\\ 
Cantoblanco E-28049 Madrid, Spain}

\author{K. MALTMAN\footnote{Permanent address: Math and Statistics, 
York University, Toronto, ON Canada M3J 1P3}{}\ and 
J.M. ZANOTTI}

\address{CSSM, University of Adelaide,\\Adelaide, SA 5005, Australia}

\maketitle

\begin{history}
\received{Day Month Year}
\revised{Day Month Year}
\end{history}

\begin{abstract}
Recent analyses of flavor-breaking 
hadronic-$\tau$-decay-based sum rules produce values of $\vert V_{us}\vert$ 
$\sim 3\sigma$ low compared to 3-family unitarity expectations. An 
unresolved systematic issue is the significant variation in 
$\vert V_{us}\vert$ produced by different prescriptions 
for treating the slowly converging $D=2$ OPE series. We investigate
the reliability of these prescriptions using lattice data 
for various flavor-breaking correlators and show the fixed-scale
prescription is clearly preferred. Preliminary updates 
of the conventional $\tau$-based, and related 
mixed $\tau$-electroproduction-data-based, sum rule analyses incorporating
B-factory results for low-multiplicity strange $\tau$ decay mode
distributions are then performed. Use of the preferred FOPT $D=2$ OPE 
prescription is shown to significantly reduce the discrepancy between
3-family unitarity expectations and the sum rule results.
\end{abstract}
\vskip .15in
The conventional inclusive hadronic $\tau$ decay determination
of $\vert V_{us}\vert$~\cite{gamizetalvus} is obtained by applying 
the finite energy sum rule (FESR) relation, involving polynomial weight
$w(s)$ and kinematic-singularity-free correlator
$\Pi (s)$ with spectral function $\rho (s)$,
\begin{equation}
\int_0^{s_0}w(s) \rho(s)\, ds\, =\, -{\frac{1}{2\pi i}}\oint_{\vert
s\vert =s_0}w(s) \Pi (s)\, ds\ ,
\label{basicfesr}
\end{equation}
to the flavor-breaking (FB) difference 
$\Delta\Pi_\tau \, \equiv\,
\left[ \Pi_{V+A;ud}^{(0+1)}\, -\, \Pi_{V+A;us}^{(0+1)}\right]$,
where $\Pi^{(J)}_{V/A;ij}(s)$ are the spin $J=0,1$ components of the 
flavor $ij$, vector (V) or axial vector (A) current-current 2-point
functions. The spectral functions, $\rho^{(0+1)}_{V/A;ij}$, hence also
$\Delta\rho_\tau$, are related to the normalized differential 
decay distributions, $dR_{V/A;ij}/ds$, of flavor $ij$ V- or 
A-current-induced $\tau$ decay widths,
$R_{V/A;ij}\, \equiv\, \Gamma [\tau^- \rightarrow \nu_\tau 
\, {\rm hadrons}_{V/A;ij}\, (\gamma )]/ \Gamma [\tau^- \rightarrow
\nu_\tau e^- {\bar \nu}_e (\gamma)]$, by
\begin{equation}
dR_{V/A;ij}/ds\, =\, 12\pi^2\vert V_{ij}\vert^2 S_{EW}\left[
w_\tau (y_\tau )\rho^{(0+1)}_{V/A;ij}(s)-w_L(y_\tau )\rho^{(0)}(s)\right]\,
/m_\tau^2\ ,
\label{dRds}\end{equation}
with $y_\tau =s/m_\tau^2$, $V_{ij}$ the $ij$ CKM matrix element, 
$w_\tau (y)=(1-y)^2(1+2y)$, $w_L(y)=y(1-y)^2$, and $S_{EW}$ a short-distance 
electroweak correction factor. The $J=0$ (longitudinal) 
contributions in (\ref{dRds}) are well known phenomenologically and, due 
to problems with the corresponding $D=2$ OPE series, usually subtracted 
from $dR/ds$~\cite{gamizetalvus,kmetalvus}. The subtracted result, 
$dR^{(0+1)}_{V/A;ij}/ds$, allows the construction of $J=0+1$ reweighted 
analogues, 
$R^w_{V+A;ij}(s_0)\, =\, \int_0^{s_0}ds\, \left[ w(s)/w_\tau (y_\tau )\right]
\, dR^{(0+1)}_{V+A;ij}(s)/ds$, for any $w(s)$ and 
$s_0<m_\tau^2$. Defining 
$\delta R^w_{V+A}(s_0)\, =\,
\left[ R^w_{V+A;ud}(s_0)/\vert V_{ud}\vert^2\right]
\, -\, \left[ R^w_{V+A;us}(s_0)/\vert V_{us}\vert^2\right]$, one has,
for $s_0$ large enough to allow use of the OPE 
on the RHS of (\ref{basicfesr}),~\cite{gamizetalvus}
\begin{equation}
\vert V_{us}\vert \, =\, \sqrt{R^w_{V+A;us}(s_0)/\left[
{\frac{R^w_{V+A;ud}(s_0)}{\vert V_{ud}\vert^2}}
\, -\, \delta R^{w,OPE}_{V+A}(s_0)\right]}\ .
\label{tauvussolution}\end{equation}
This relation has usually been employed in un-reweighted form, with 
$w=w_\tau$, and the single value $s_0=m_\tau^2$~\cite{gamizetalvus}. This 
has the advantage that $R^{w_\tau}_{V+A;ud,us}(m_\tau^2)$ is determinable 
from branching fraction information alone, but the disadvantage of precluding 
tests of the $s_0$- and $w(s)$-independence of the analysis, which could 
otherwise be used to investigate potential systematic uncertainties (in 
particular, those associated with the treatment of OPE contributions). Such 
self-consistency tests were carried out in 
Refs.~\refcite{kmetalvus,kmfoptvus1,kmfoptvus2}, and 
non-trivial $w(s)$- and $s_0$-dependences observed, 
suggesting shortcomings in the experimental data and/or OPE representation. 

The most obvious potential OPE problem lies in the rather slow convergence 
of the $D=2$ OPE series. In terms of
the running $\overline{MS}$ quantities $m_s(Q^2)$ and $\bar{a}
\equiv \alpha_s(Q^2)/\pi$, the $D=2$ series, which is known to
4-loops, is given by
\begin{eqnarray}
&&\left[\Delta\Pi_\tau (Q^2)\right]^{OPE}_{D=2}\, =\, {\frac{3}{2\pi^2}}\,
{\frac{m_s^2(Q^2)}{Q^2}} \sum_{k=0}c^\tau_k \bar{a}^k
\label{d2form}\end{eqnarray}
with $c^\tau_k\, =\, 1$, $7/3$, $19.93$, $208.75$ for 
$k=0\cdots 3$~\cite{bckd2ope}. Since
$\bar{a}(m_\tau^2)\simeq 0.10$, $c^\tau_3\bar{a}^3>
c^\tau_2\bar{a}^2$ at the spacelike point on the contour 
for all $s_0\le m_\tau^2$. The problematic convergence complicates the 
assessment of $D=2$ truncation errors, and manifests itself, e.g., in 
the $\sim 0.0020$ difference in $\vert V_{us}\vert$ values obtained using 
two alternate (CIPT or FOPT) versions of the 4-loop-truncated, 
$w_\tau$-weighted series.

An alternate determination employs the FB 
combination $\Delta\Pi_{\tau -EM}\, \equiv\, 9\Pi_{EM} - 5\Pi_{ud;V}^{(0+1)} 
+ \Pi_{ud;A}^{(0+1)} - \Pi^{(0+1)}_{us;V+A}$ in place of 
$\Delta\Pi_\tau$~\cite{kmtauem08}. Inclusive electroproduction 
cross-sections fix the electromagnetic (EM) spectral function. 
By construction, the $\Delta\Pi_{\tau -EM}$ $D=2$ series 
is strongly suppressed, having the form (\ref{d2form}),
with $c_k^\tau\rightarrow c_k^{\tau -EM}\, =\, 0$, 
$-1/3$, $-4.384$, $-44.943$ for $k=0\cdots 3$. The $D=4$ series is also 
strongly suppressed. OPE contributions to $\Delta\Pi_{\tau -EM}$ FESRs, 
hence also estimated OPE errors, are thus very small~\cite{kmtauem08}, and 
the resulting $\vert V_{us}\vert$ errors essentially entirely experimental. 
A check of this predicted suppression is thus of interest.

We investigate the relative merits of the fixed-scale (FOPT-like) and 
local-scale ($\mu^2=Q^2$, i.e., CIPT-like) treatments of the 
$\Delta\Pi_\tau$ $D=2$ series, and the level of $\Delta\Pi_{\tau -EM}$ 
suppression, by comparing OPE expectations and lattice data 
for the two correlator combinations over a range of Euclidean $Q^2$. 
Five RBC/UKQCD domain wall fermion ensembles are employed, three, with 
$m_\pi =293,\, 349,\, 399\ MeV$, having $1/a=2.31\ GeV$~\cite{ainv228},
and two, with $m_\pi =171,\, 248\ MeV$, having $1/a=1.37\ GeV$~\cite{ainv137}. 
For technical reasons, conserved-local versions of the flavor $us$ 2-point 
functions are numerically challenging and hence, for $\Delta\Pi_\tau$, 
local-local versions are used. To check that this does not produce residual 
lattice artifacts which would impact our conclusions,
we have also performed the OPE-lattice comparison, using conserved-local 
data, for the alternate flavor-diagonal FB combination 
$\Delta\Pi_{diag}\equiv \Pi_{V;\ell\ell}-\Pi_{V;ss}$, 
whose $D=2$ series is very similar to that of $\Delta\Pi_\tau$ 
($c^\tau_k\rightarrow c_k^{diag}\, =\, 1$, $8/3$, $24.32$, $253.69$ 
for $k=0\cdots 3$ in (\ref{d2form})). The results confirm those of 
the local-local study.

\begin{center}
\begin{figure}[h]
  \begin{minipage}[h]{0.47\linewidth}
\rotatebox{270}{\mbox{
\includegraphics[width=0.92\textwidth]
{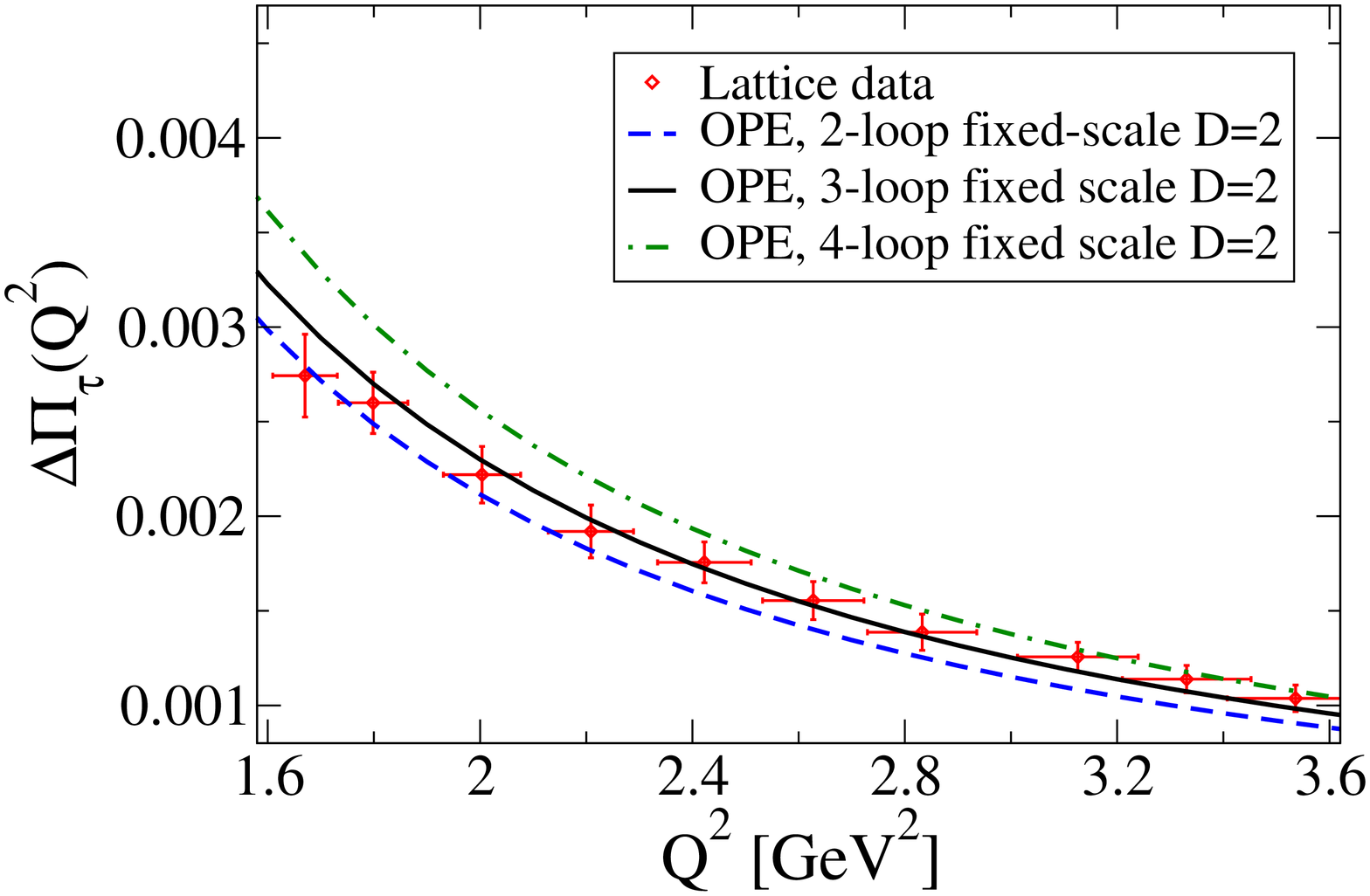}
}}
  \end{minipage}
\hfill
  \begin{minipage}[h]{0.47\linewidth}
\rotatebox{270}{\mbox{
\includegraphics[width=0.92\textwidth]
{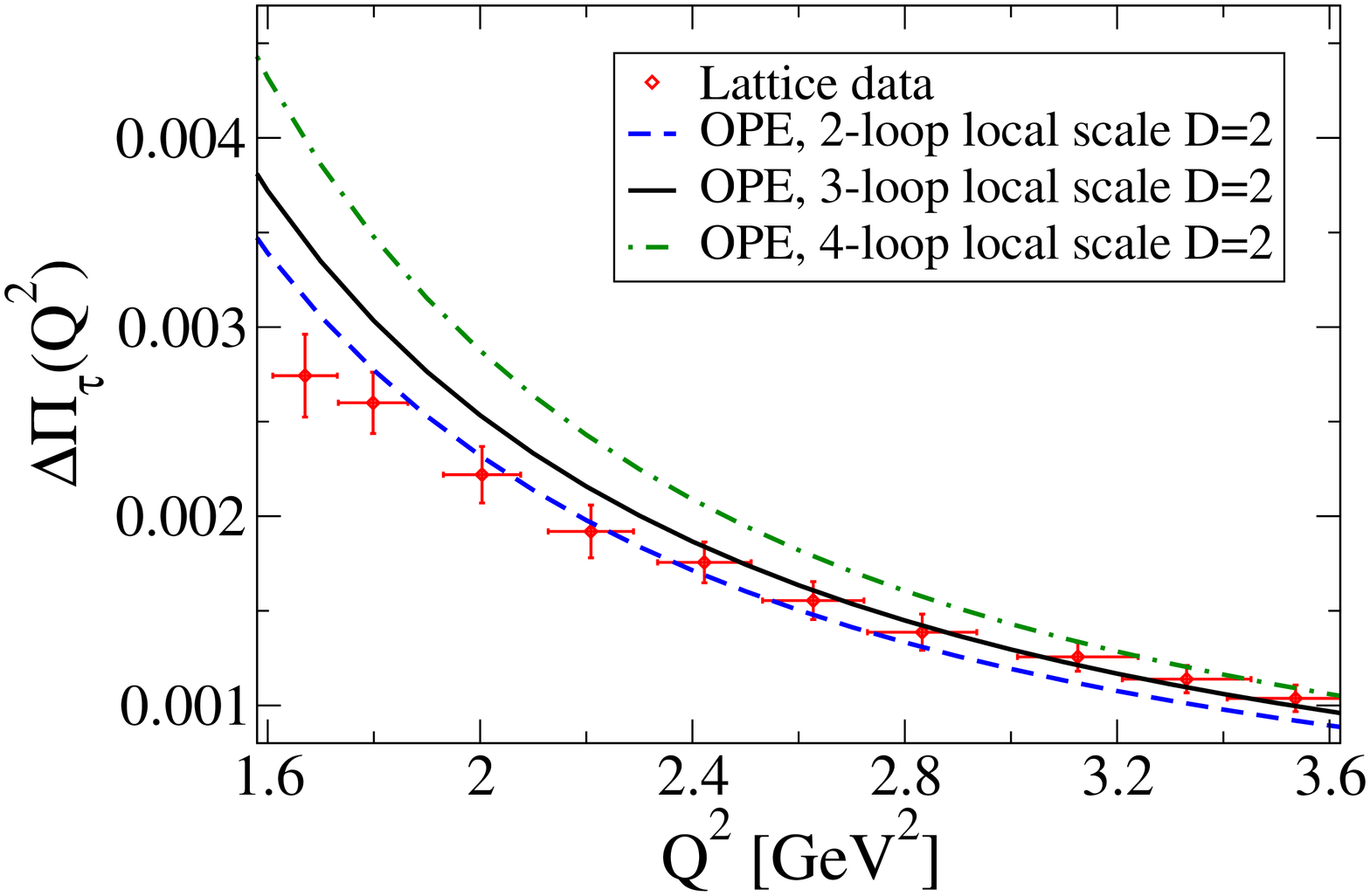}
}}
\end{minipage}
\vspace*{8pt}
\caption{OPE and lattice $\Delta\Pi_\tau$ data,
$1/a = 2.31$ GeV, $m_\pi =293$ MeV ensemble, $O(\bar{a}^{1,2,3})$
$D=2$ OPE truncation, fixed-scale (left panel) or local-scale 
(right panel) $D=2$ prescription \label{figtaufb}}
\end{figure}
\end{center}

Representative OPE-lattice data comparisons for $\Delta\Pi_\tau$ 
are shown, for the $1/a=2.13$ GeV, $m_\pi =293$ MeV ensemble, 
in Fig.~\ref{figtaufb}. The left (right) panel comparison
employs the fixed-scale (local-scale) prescription for the $D=2$ 
OPE series. The fixed-scale versions match much better the 
$Q^2$ dependence of the lattice results, with the 3-loop-truncated 
version thereof best matching the overall normalization.
\vskip -0.25in
\vskip -.15in

The comparison of lattice data for $\Delta\Pi_\tau$ and
$\Delta\Pi_{\tau -EM}$ confirms the very strong suppression of
$\Delta\Pi_{\tau -EM}$~\cite{kmfoptvus2} (see Ref.~\refcite{kmfoptvus2} 
for the relevant figure).
\begin{center}
\begin{figure}[h]
  \begin{minipage}[h]{0.47\linewidth}
\rotatebox{270}{\mbox{
\includegraphics[width=0.92\textwidth]
{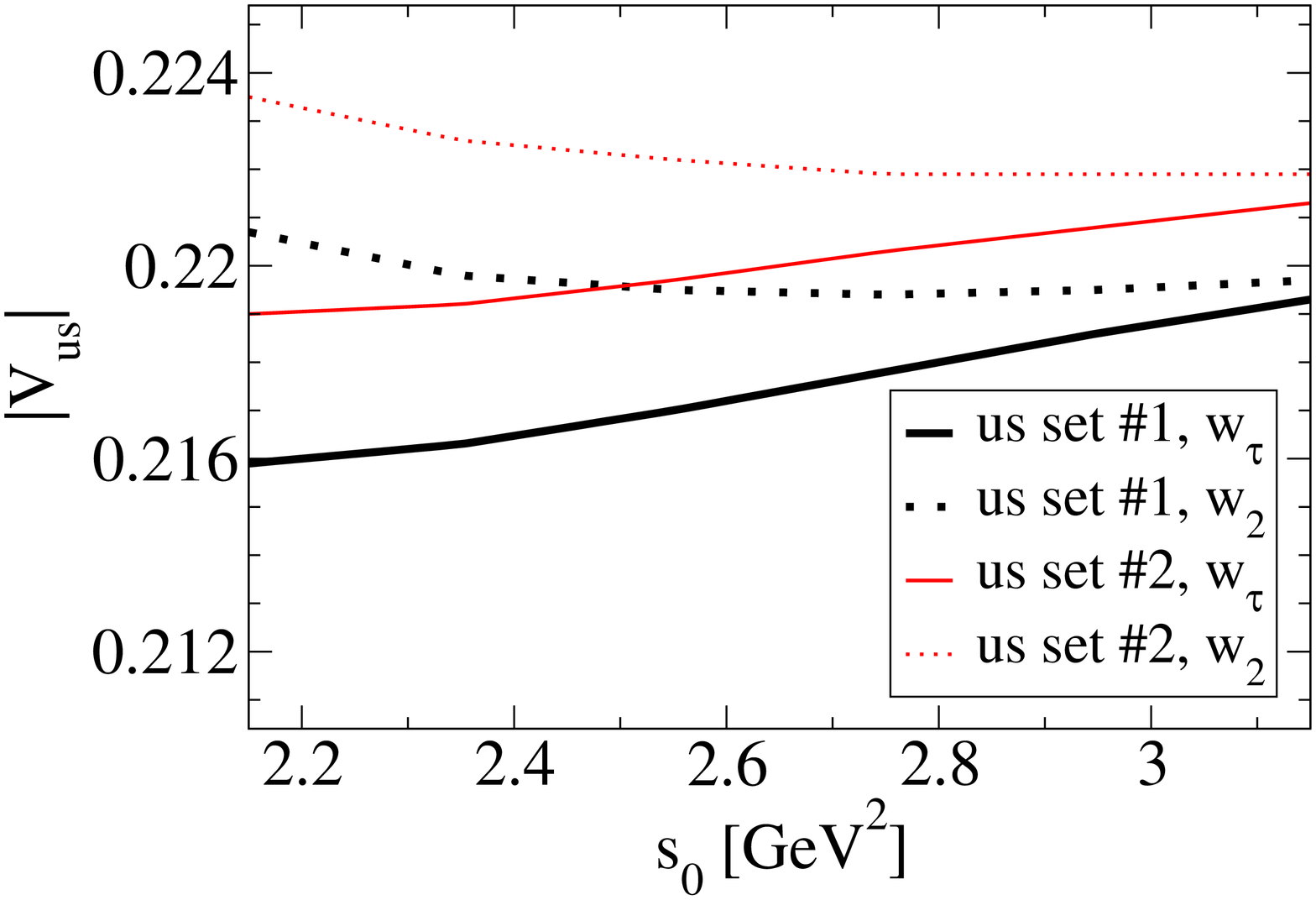}
}}
  \end{minipage}
\hfill
  \begin{minipage}[h]{0.47\linewidth}
\rotatebox{270}{\mbox{
\includegraphics[width=0.92\textwidth]
{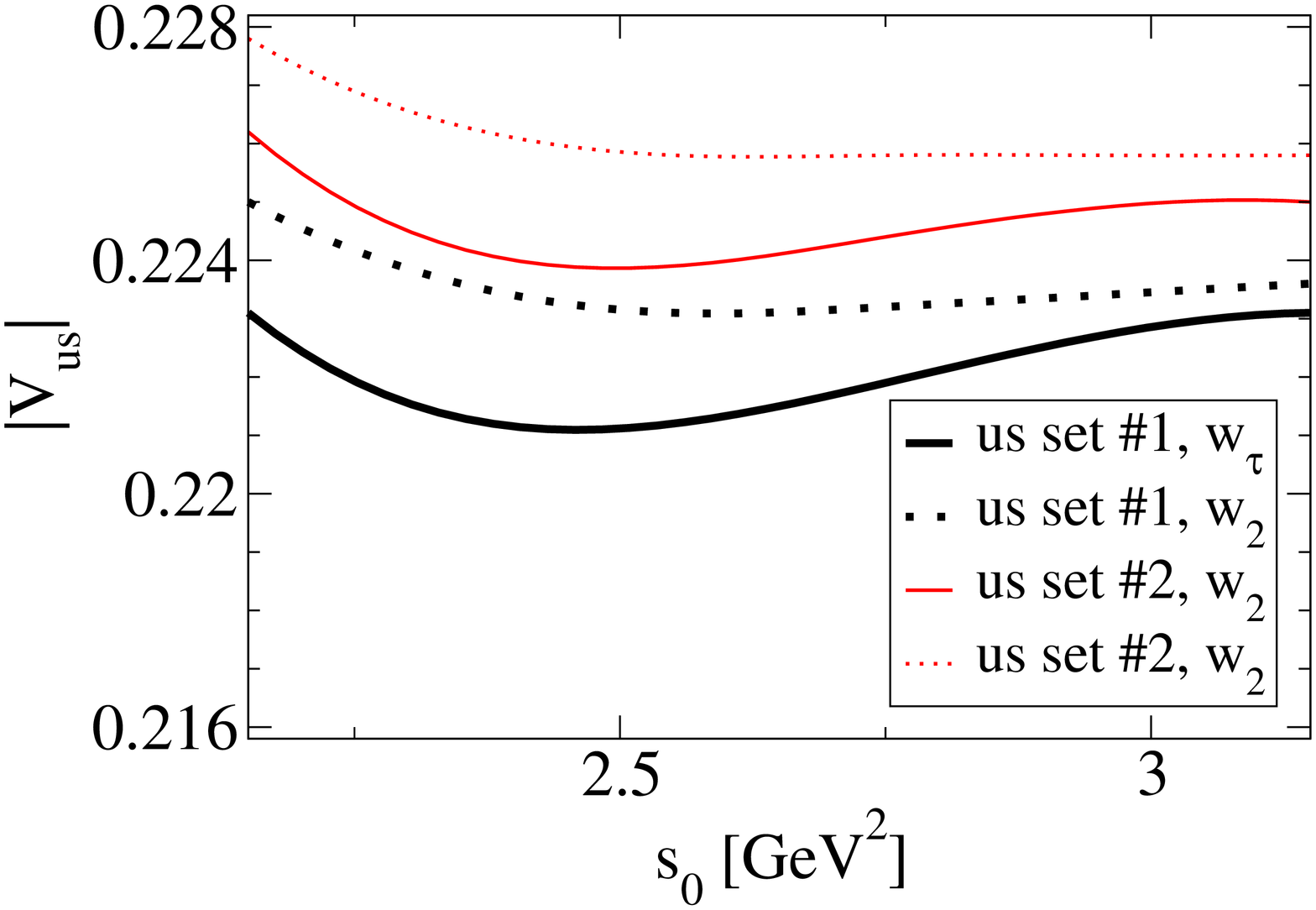}
}}
\end{minipage}
\vspace*{8pt}
\caption{$\vert V_{us}\vert$ from 
preliminary updates of the FB $\Delta\Pi_\tau$ and 
$\Delta\Pi_{\tau -EM}$ FESRs \label{figtaufbtauemvus}}
\end{figure}
\end{center}
\vskip -.15in

We turn to preliminary updates of the
$\vert V_{us}\vert$ analyses. For the $D=2$ OPE series, we employ 
the 3-loop-truncated FOPT prescription favored by lattice data, and
for the $ud$ spectral integrals, OPAL data~\cite{opalud}, as updated in
Ref.~\refcite{dv7}. For the $us$ spectral integrals, recent B-factory 
results are used for the $K\pi$~\cite{babarbellekpi},
$K^-\pi^-\pi^+$~\cite{babarkpipiallchg} and 
$K_s\pi^-\pi^0$~\cite{bellekspipi} exclusive mode distributions, and ALEPH
results~\cite{alephus}, updated for current branching fractions (BFs),
for all other modes. Contributions from the latter lie higher 
in the spectrum, and have much larger errors. The B-factory
distributions are unit normalized, and also require current
BFs for their overall scales. We work with BFs obtained in a 
$\pi_{\mu 2}$, $K_{\mu 2}$-constrained HFAG fit, supplemented by the 
update to $B[\tau^-\rightarrow K_s^0\pi^-\pi^0\nu_\tau ]$ produced by 
the recent Belle result~\cite{bellekspipi}. Other non-trivial shifts 
in the $us$ BFs also remain possible.
To illustrate the changes to $\vert V_{us}\vert$ that could result,
we consider also an alternate set of $us$ BFs with the recent larger, 
but not yet finalized, BaBar results~\cite{adametz} for 
$B[\tau^-\rightarrow K^-\, n\, \pi^0\nu_\tau ]$, $n\leq 3$, 
used in place of those of the HFAG fit. The first set of $us$ BFs
is labelled ``$us$ BF set \#1'' below, the second, alternate set
``$us$ BF set \#2''. Changes to the $us$ BFs alter the inclusive 
$us$ spectral distribution, and hence can affect both the magnitude 
of $\vert V_{us}\vert$ and the $s_0$-dependence of the results. 
The significantly larger preliminary BaBar $K^-\pi^0$ BF is 
particularly relevant for the FB FESRs considered here, which 
weight more strongly the low-$s$ part of the spectrum. We consider FESRs 
employing the weights $w_\tau$ and $w_2(y)=(1-y)^2$. $w_2$ weights less
strongly the higher-$s$, large-error region of the $us$ spectral distribution.
Differences between results obtained using the two different 
weights can thus point to issues with the $us$ spectral distribution. 

$\vert V_{us}\vert$ results obtained from the $w_\tau$ and $w_2$
versions of the $\Delta\Pi_\tau$ FESR are shown, as a function of
$s_0$, and also the choice of the input $us$ BF set, in the left
panel of Fig.~\ref{figtaufbtauemvus}. Similar results for the
$\Delta\Pi_{\tau -EM}$ FESR are shown in the right panel. 
$w_2$ results, which are less sensitive to the large-error high-$s$ region,
show better $s_0$-stability in both cases. For $w_\tau$,
$s_0$-stability is also better for the $\Delta\Pi_{\tau -EM}$ case,
where OPE contributions are suppressed. The convergence of $w_\tau$ results
to the more stable $w_2$ ones as $s_0\rightarrow m_\tau^2$, seen for
both the $\Delta\Pi_\tau$ and $\Delta\Pi_{\tau -EM}$ FESRs, suggests
the possibility of residual OPE problems in the $w_\tau$ case, where
cancellations on the contour play a larger role. Finally we note that
results obtained using the FOPT prescription preferred by the
lattice data agree better with 3-family unitarity expectations 
than do those (not shown here) obtained using CIPT, as do those obtained 
using $us$ BF set \#2 in place of $us$ BF set \#1. More details of these 
analyses will be presented elsewhere.

We close by stressing the preference for FOPT over CIPT for
the $D=2$ OPE series. The prescription which underlies CIPT (of summing
logarithmic terms to all orders while truncating the series
of non-logarithmic terms), though plausible, is motivated by
heuristic arguments not generally valid for divergent series~\cite{bj08}, 
and performs poorly when tested against lattice data for the FB
correlators. 

\section*{Acknowledgments}
Computations were performed using the STFC's DiRAC
facilities at Swansea and Edinburgh. PAB, LDD, and RJH were supported
by an STFC Consolidated Grant, and by the EU under Grant Agreement
PITN-GA-2009-238353 (ITN STRONGnet); EK by the Comunidad
Aut\`onoma de Madrid under the program HEPHACOS S2009/ESP-1473 and the
EU under Grant Agreement PITN-GA-2009-238353 (ITN STRONGnet); KM
by NSERC (Canada); and JMZ by the Australian Research Council grant 
FT100100005.




\end{document}